\documentclass[fleqn,twoside]{article}
\usepackage{espcrc2}

\usepackage{graphicx}
\usepackage[figuresright]{rotating}

\newcommand{\MSb}{$\overline{\rm MS}$ }
\newcommand{\ep}{\varepsilon}

\title{Steps towards full two-loop calculations
      for 2 fermion to 2 fermion processes: 
      running versus pole masses schemes}

\author{F. Jegerlehner \address[MCSD]{DESY, Platanenallee 6, 
                        D-15738, Zeuthen, Germany} 
\thanks{~E-mail: fred.jegerlehner@desy.de}
and 
    M.Yu Kalmykov\addressmark[MCSD] \thanks{
Supported by DFG under Contract SFB/TR~9-03.} 
\thanks{
On leave from BLTP, JINR, 141980 Dubna, Russia}
}
       
\begin{document}

\begin{abstract}
Recent progress in the calculation of the two-loop on-shell mass
counterterms within the electroweak Standard Model (SM) for the
massive particles are discussed.  We are in progress of developing a
package for full two-loop SM calculations of 2 $\to$ 2 fermion
processes, with emphasis on the analytical approach where
feasible. The complete two-loop on-shell renormalization is
implemented. Substantial progress has been made in calculating the
master integrals. We are able to compute in an efficient and stable
manner up to a few thousands of diagrams of very complex mass
structure.
\vspace{1pc}
\end{abstract}

\maketitle
\section{Introduction}
Even for the simplest physical processes like $e^+e^- \to f\bar{f}$
complete two-loop electroweak SM calculations still are not available,
maninly because of the enourmous complexity of such calculations. An
exception is the $\mu$--decay rate~\cite{Awramik:2002wn}, which is
simpler due to the fact that it is a static quantity. Besides the
large number of diagrams encountered, the difficulties start at the
level of individual Feymnan integrals and increase substantially when
going from propagators to vertices or box contributions.  One key
problem at the beginning is the renormalization and the calculation of
the necessary counterterms. In the QED like on-shell renormalization
scheme the basic input parameters for electroweak higher order
calculations are the fine structure constant $\alpha=\frac{e^2}{4\pi}$
and all the physical particle masses. In this scheme the whole
renormalization program only requires the calculation of selfenergy
diagrams and tadpoles. While the calculation of the counterterm for
the fine structure constant is relatively easy (at zero
momentum)~\cite{Degrassi:2003rw}, the on-shell mass counterterms are
much more involved. In this note we therefore focus on aspects of
calculating the latter, which yields the relation between bare,
$\overline{\rm MS}$ and on-shell (pole) masses (two-loop
renormalization constant in on-shell scheme). Many details on this
program may be found in the original
publications~\cite{poleI,pole-top} and will not be repeated here. We
rather concentrate on some controversal points concerning the
definition of $\overline{\rm MS}$ masses in electroweak theory
(Sec.~2). We also give some more specific information concerning our
numerical approach to the calculation of huge sets of diagrams
(Sec.~3).

\section{\protect $\overline{\rm MS}$-masses of particles in SM and renormalization~~group~~equations}

The two-loop calculation of pole masses of the gauge bosons in the SM
has been discussed in~\cite{poleI,pole-top}. In terms of the
transversal self-energy function $\Pi(p^2,m^2,\cdots)$, by expansion
about $p^2=-m^2$, will get the 
two loop solution
\begin{eqnarray}
&& \hspace{-5mm}
s_P = m^2 - \Pi^{(1)} - \Pi^{(2)}
- \Pi^{(1)} \Pi^{(1)}{}' \;,
\label{polemass}
\end{eqnarray}
for the location of the pole $s_P$. $\Pi^{(L)}$ is the bare or
$\overline{\rm MS}$-renormalized $L$-loop contribution to $\Pi$, the
prime denotes the derivative with respect to $p^2$. One of the
remarkable properties of (\ref{polemass}) is that the complex pole is
represented by the self-energy and its derivative at momentum equal to
the bare or $\overline{\rm MS}$ mass which, by definition, are real
parameters. The standard parametrization of the pole is $s_{P,a} =
M^2_a - i M_a \Gamma_a$, where $M_a$ is the pole mass and $\Gamma_a$
is the width of particle $a$.

For calculations of electroweak corrections in the SM two
renormalization schemes are commonly accepted: the on-shell and the
$\overline{\rm MS}$ scheme.  It is well understood, that in the
on-shell scheme all momentum independent diagrams, in particular the
tadpoles, can be omitted (the on-shell scheme is a particular case of
a momentum subtraction scheme: finite parts are fixed by subtraction
of the propagators at $p^2=-s_P$). The set of diagrams contributing to
the $\overline{\rm MS}$ mass are not unambiguously defined. Let us
first express the pole (\ref{polemass}) in terms of the bare amplitude
in a manifestly gauge invariant manner. This requires to include the
Higgs tadpole contribution~\cite{FJ}.  Only this complete
gauge invariant bare amplitude should be utilized as a starting point
to set up $\overline{\rm MS}$ renormalization. At
the two-loop level $\overline{\rm MS}$ renormalization can be written
as\\[-3mm]
\begin{eqnarray}
&& \hspace{-5mm}
s_{P} =  m_{0}^2 \!-\! \Pi^{(1)}_{0} 
\!-\! \Pi_{0}^{(2)} \!-\! \Pi^{(1)}_{0} \Pi_{0}^{(1)}{}'
\nonumber\\ && \hspace{-5mm}
\!-\! \Biggl[ \sum\limits_j (\delta m^2_{j,0})^{(1)} \frac{\partial}{\partial m_{j,0}^2} 
\!+\! \sum\limits_j (\delta g_{j,0})^{(1)} \frac{\partial}{\partial g_{j,0}}
\Biggr] \Pi_{0}^{(1)} 
\nonumber\\ && \hspace{-5mm}
= m^2_a 
\!-\! \left \{ \! \Pi_a^{(1)} \! \right\}_{\overline{\rm MS}}
\!-\! \left \{ \! \Pi_a^{(2)} \!+\! \Pi_a^{(1)} \Pi_a^{(1)}{}' \! \right\}_{\overline{\rm MS}}
\label{mct}
\end{eqnarray}
where the sum runs over all species of particles, $g_j = \alpha$,
$g_s$, $(\delta g_{j,0})^{(1)}$ and $(\delta m^2_{j,0})^{(1)}$ are the
one-loop counterterms for the charges and physical masses in the
{\MSb}-scheme and after differentiation we put all parameters equal to
their on-shell values. The derivatives in Eq.~(\ref{mct}) correspond
to the subtraction of sub-divergencies.  The genuine two-loop mass
counterterm comes from the shift of the $m_{0}^2$ term.
The relation between bare- and \MSb-masses has the
form\\[-3mm]
\begin{equation}
m_{a,0}^2 =  m_a^2(\mu)\:
( 1 + \sum_{k=1} Z_a^{(k)} \ep^{-k} )\;. 
\label{singular}
\end{equation}
\noindent
To renormalize the pole mass at the two-loop level requires to
calculate the one-loop renormalization constants for all physical
parameters (charge and masses), and the two-loop renormalization
constant only for the mass itself. Not needed are the wave-function
renormalization or ghost (unphysical) sector renormalizations.  After
UV-renormalization the pole is represented in terms of finite
amplitudes.  Now, expression (\ref{mct}) connects the pole $s_P$ with
the $\overline{\rm MS}$ parameters: masses and charges. This
expression can be inverted and solved iteratively.  The solution to
two-loop reads\\[-3mm]
\begin{eqnarray}
&& \hspace{-7mm}
m_a^2 \!=\! M_a^2
\!+\! {\rm Re} \left \{ \! \Pi_a^{(1)} \! \right\}_{\overline{\rm MS}}
\!+\! {\rm Re} \left \{ \! \Pi_a^{(2)} \!+\! \Pi_a^{(1)} \Pi_a^{(1)}{}' \! \right\}_{\overline{\rm MS}}
\nonumber\\ && \hspace{-7mm}
\!+\! \left[ \! 
(\Delta e)^{(1)} \frac{\partial}{\partial e} 
\!+\! 
\sum\limits_j (\Delta m^2_j)^{(1)} \frac{\partial}{\partial m_j^2} 
\! \right]
{\rm Re} \left \{ \Pi_a^{(1)} \right\}_{\overline{\rm MS}}  
\label{reverse}
\end{eqnarray}
\noindent
where the sum runs over all species of particles $j=Z,\,W,\,H, \, t$,
$ (\Delta m^2_j)^{(1)}= {\rm Re} \left \{ \Pi_j
\right\}_{\overline{\rm MS}} \;, $ and the transition from the \MSb to
the on-shell scheme for the electric charge~\cite{FJ03} is also
included.  The mass on the l.h.s. of this expression we call the
${\overline{\rm MS}}$-mass of particle.  It should be noted, that in
this definition the tadpole contribution does not cancel, so that
higher powers of the Higgs and the top-quark mass show up at higher
orders.  In particular, at two-loops, the purely bosonic diagrams
generate $m_H^4/m_V^4$ terms and the third fermion family gives rise to
the appearance of $m_t^6/(m_H^2 m_V^4)$ power corrections.  For the
${\overline{\rm MS}}$-masses, defined in this way, the following
properties are valid:

\noindent
${\bf 1.}$
The UV counter-terms satisfy relations connecting the higher order poles
with the lower order ones:
\begin{eqnarray}
&& \hspace{-5mm}
\biggl(
\gamma_a \!+\!\sum_j \beta_{g_j} \frac{\partial}{\partial g_j }
\!+\! \sum_i \gamma_i m_i^2 \frac{\partial}{\partial m_i^2} \biggr) Z_a^{(n)}
\nonumber \\ && \hspace{-5mm}
=  \frac{1}{2} \sum_j g_j \frac{\partial}{\partial g_j } Z_a^{(n+1)} \;, 
\end{eqnarray}
\noindent
where we adopt the following definitions for the RG functions: for all
dimensionless coupling constants, like $g,g',g_s,e,\lambda, y_t$, the
$\beta$-function is given by 
$\mu^2 \frac{\partial}{\partial \mu^2} g = \beta_g$ and for all mass
parameters (a mass or the Higgs v.e.v. $v$) the anomalous dimension
$\gamma_{m^2}$ is given by $\mu^2
\frac{\partial}{\partial \mu^2} \ln m^2 = \gamma_{m^2}.$ 

\noindent
${\bf 2.}$
Using the fact that $s_P$ is RG-invariant:  
$\mu^2 \frac{d}{d \mu^2} s_P \!\equiv \!0$, 
we are able to calculate the anomalous dimension of the
masses from our finite results (\ref{reverse})
or from the UV counterterms (\ref{singular})\\[-3mm]
$$
\gamma_a = \sum_j \frac{1}{2} g_j \frac{\partial}{\partial g_j } Z_a^{(1)}
\;, (j=g,g_s).$$

\noindent
${\bf 3.}$
All tree level relations between masses of any particles
and parameters of the unbroken Lagrangian are RG invariant. This means, 
in particular, that the RG equation for the vacuum expectation value $v$ is
given by $\gamma_{v^2} \equiv \gamma_{m^2} - \beta_\lambda/\lambda$, 
where $m^2$ and $\lambda$ are the parameters of the symmetric scalar potential.
This fact allow to get anomalous dimension of the masses via the relations \cite{RG:our}
\vspace*{-4mm}
\begin{eqnarray}
&& \hspace{-7mm}
\gamma_W  =  \gamma_{m^2} - \frac{\beta_\lambda}{\lambda} + 2 \frac{\beta_g}{g} \;, 
\nonumber \\ && \hspace{-7mm}
\gamma_Z = \gamma_{m^2} \!-\! \frac{\beta_\lambda}{\lambda} 
\!+\! 2  \left( c_W \frac{\beta_g}{g} \!+\! s_W \frac{\beta_{g'}}{g'} \right),\nonumber \\
\nonumber \\ && \hspace{-7mm}
\gamma_t  =  \gamma_{m^2} - \frac{\beta_\lambda}{\lambda} + \frac{\beta_{y_t}}{y_t} \;,
\quad
\gamma_H  =  \gamma_{m^2} \;,
\label{rg2}
\end{eqnarray}
where $s_W (c_W) $ are the $\sin$ ($\cos$) of the weak mixing angle
and the 2-loop RG functions $\beta_g, \beta_{g'}, \beta_\lambda,
\gamma_{m^2}, \beta_{y_t}$ are calculated in the unbroken phase \cite{RG}.

\noindent
The RG invariance of the pole positions $s_{P}$ allow us 
to factorize explicitely the RG logarithms \\[-3mm] 
\begin{eqnarray}
&& \hspace{-5mm}
M_a^2 = 
m_a^2 \!-\! \sum_j g_j^2 \left ( m_a^2 \gamma_j^{(a)} L_b \!-\! X^{(a)}_{j}  \right)
\nonumber \\ && \hspace{-5mm}
\!+\! \sum_{i,j} g_i^2 g_j^2
\left[ 
m_a^2 \left( C_{i,j;a}^{(2,2)} L_b^2 \!+\! C_{i,j;a}^{(2,1)} L_b \right) 
\!+\! X^{(a)}_{i,j} \right],   
\nonumber 
\end{eqnarray}
where $L_b = \ln \mu^2/m_b^2$, $ C_{i,j;a}^{(m,n)} =  C_{j,i;a}^{(m,n)}$ 
and 
\begin{eqnarray}
&& \hspace{-5mm}
\mu^2 \frac{\partial}{\partial \mu^2} \ln m_k^2 \!=\! 
\gamma^{(k)} \!=\!  
\sum_j g_j^2 \gamma^{(k)}_j \!+\! \sum_{i,j} g_i^2 g_j^2 \gamma^{(k)}_{i,j}
\nonumber \\ && \hspace{-5mm}
\mu^2 \frac{\partial}{\partial \mu^2} g_k = 
\sum_j g_j^3 \beta_j + \sum_{i,j} g_i^2 g_j^2 \beta_{i,j} \;, 
\nonumber \\ && \hspace{-5mm}
C_{i,j;a}^{(2,1)} \!=\! \gamma^{(a)}_{i,j}
\!+\! \frac{1}{2} \!\left(\! \gamma_i^{(a)} \gamma_j^{(b)} \!+\! \gamma_i^{(b)}  \gamma_j^{(a)}  
                  \!\right)\!
\!+\! 2 \beta_j \delta_{i,j} \frac{X^{(a)}_j}{m_a^2}
\nonumber \\ && \hspace{-5mm}
\!+\!  \frac{1}{2} \sum_k \frac{m_k^2}{m_a^2}
\left(  
  \gamma^{(k)}_{i} \frac{\partial}{ \partial m_k^2}X^{(a)}_j  
\!+\! \gamma^{(k)}_{j} \frac{\partial}{ \partial m_k^2}X^{(a)}_i  
\right) \;, 
\nonumber \\ && \hspace{-5mm}
2 C_{i,j;a}^{(2,2)} \!=\! 2 \beta_j \gamma_j^{(a)} \delta_{i,j}
\!+\! \gamma_i^{(a)} \gamma_j^{(a)}
\nonumber \\ && \hspace{-5mm}
\!+\! \frac{1}{2} \sum_k \left(
  \gamma^{(k)}_{i} m_k^2 \frac{\partial}{ \partial m_k^2} \gamma^{(a)}_j  
\!+\! \gamma^{(k)}_{j} m_k^2 \frac{\partial}{ \partial m_k^2} \gamma^{(a)}_i  
\right)
\;,
\nonumber 
\end{eqnarray}
\noindent
where 
$X_j^{(a)}$ and their derivatives can be extracted 
from Appendixes of~\cite{FJ,poleI,pole-top}.

\noindent
Crucial point of our definition of the {\MSb}-mass (\ref{reverse}) is
the gauge invariant construction (\ref{polemass}) for the
pole in terms of the unrenormalized, bare diagrams. It can be done
only after inclusion of the Higgs tadpole contribution. Another
important ingredient are the Ward identities.

\noindent
${\bf A.}$ The inclusion of the tadpoles is necessary to ensure, that
the physical Higgs field has zero vacuum expectation value in each
order of the loop expansion.

\noindent
${\bf B.}$ It is well know, that in order to preserve the Ward
identities for the longitudinal part of the gauge boson propagator it
is necessary to add the tadpole contribution, which is equal to the
propagator of the would-be-Goldstone bosons at zero momentum transfer.
In particular, at the two-loop level, the photon would aquire a mass
if the tadpole contribution would be omitted.

\noindent
Our RG equations (\ref{rg2}) for the v.e.v. $v$ and the particle
masses $m$ are different from the ones obtained in the effective
potential approach~\cite{RG-SM}. A comparison of predictions based on
these two approaches have been recently performed
in~\cite{RG-HighEnergy}.

These structural considerations were important to check our
calculations of the various counter\-terms.

For the calculation of the $O(\alpha \alpha_s)$ and the
$O(\alpha_s^2)$ corrections to the top-quark propagator we refer
to~\cite{pole-top} and~\cite{QCD-top}, respectively.
\section{Numerical results}
According to (\ref{polemass}) we need to calculate propagator-type
diagrams up to two loops on--shell.  To keep control of gauge
invariance we adopt the $R_\xi$ gauge with three different gauge
parameters $\xi_W,\,\xi_Z$ and $\xi_\gamma$.  For our calculation all
diagrams have been generated with the help of ${\bf
QGRAF}$~\cite{qgraf}. The C-program {\bf DIANA}~\cite{diana} then was
used together with the set of Feynman rules extracted from the package
{\bf TLAMM}~\cite{tlamm} to produce the FORM input which is suitable
for the package {\bf ONSHELL2}~\cite{onshell2} and/or for another
package based on Tarasov's recurrence relations~\cite{T97a}.  The set
of master-integrals, in the limit of massless lepton and light quarks
(see details in~\cite{RG:our}) includes diagrams with three different
massive scales. In most cases exact analytic results in terms of known
functions are not available. Thus, instead of working with the exact
formulae (which only can be evaluated numerically, at present) we
resort to some approximations, namely, we perform appropriate series
expansions in (small) mass ratios.  For diagrams with several
different masses it is possible that several small parameters are
available.  In this case we apply different asymptotic expansions
(see~\cite{2region}) one after the other. Specifically, we expand in
the gauge parameters about $\xi_i=1$, in $\sin^2
\theta_W$ and, for diagrams with Higgs or/and top-quark lines, in
$m_V^2/m_H^2$ or/and $m_V^2/m_t^2$. Numerical results are obtained
using the packages {\bf ON-SHELL2}~\cite{onshell2} and {\bf
TLAMM}~\cite{tlamm}. Since the quality of the convergence of a series
is not known a priori we have to calculate several coefficients of
each expansion (six in $\sin^2 \theta_W$ and five in mass ratios,
$m_V^2/(m_H^2,m_t^2)$ ) to keep control on the convergence. For the
one-loop diagrams and their derivatives we used the exact analytical
results, as given in~\cite{DK,poleI}. The expansion of diagrams with a
top-quark and/or a Higgs boson leads to two-loop bubble diagrams with
three massive lines (with two of the masses equal). For these master
integrals we utilized a special form of
representation~\cite{DK,poleI}. The diagrams with massless fermion
lines also demand special consideration. These diagrams develop
threshold singularities which behave like powers of $\ln \sin^2
\theta_W$. To control these terms we had to use the exact 
analytical results, which have been worked out in~\cite{poleI} using a
technique developed in~\cite{poleI,DK}. We found that after collecting
the contributions from all diagrams the threshold singularities
canceled. This is a manifestation of the infrared stability of the
pole mass of the gauge bosons. The series expansion in $\sin^2
\theta_W$ converges very well, and can be restricted to the first two
coefficients. The expansion in the remaining mass ratios require three
coefficients in order to get sufficient precision for light Higgs mass
values. The numerics has been performed in MAPLE. To get control of
the numerical stability, we run the MAPLE program with an accuracy of
100 decimals (a posteriori, as an experimental fact, we find that the
minimal accuracy is 40 decimals).


{\bf Acknowledgments.}
M.~K.'s research was supported in part by 
Heisenberg-Landau grant No.~2004 and 
by RFBR grant No.~04-02-17149.

\end{document}